\documentclass[aps,twocolumn,amsmath,amssymb,prb,superscriptaddress,reprint,showpacs]{revtex4}
\usepackage{graphicx}
\usepackage{dcolumn}
\usepackage{bm}
\usepackage{color}
\begin{document}
\title{Thermographic measurements of the spin Peltier effect in metal/yttrium-iron-garnet junction systems}
\author{S. Daimon}
\email{sdaimon@imr.tohoku.ac.jp}
\affiliation{Institute for Materials Research, Tohoku University, Sendai 980-8577, Japan}
\affiliation{WPI Advanced Institute for Materials Research, Tohoku University, Sendai 980-8577, Japan}
\author{K. Uchida}
\email{UCHIDA.Kenichi@nims.go.jp}
\affiliation{Institute for Materials Research, Tohoku University, Sendai 980-8577, Japan}
\affiliation{Center for Spintronics Research Network, Tohoku University, Sendai 980-8577, Japan}
\affiliation{National Institute for Materials Science, Tsukuba 305-0047, Japan}
\affiliation{PRESTO, Japan Science and Technology Agency, Saitama 332-0012, Japan}
\author{R. Iguchi}
\affiliation{Institute for Materials Research, Tohoku University, Sendai 980-8577, Japan}
\affiliation{National Institute for Materials Science, Tsukuba 305-0047, Japan}
\author{T. Hioki}
\affiliation{Institute for Materials Research, Tohoku University, Sendai 980-8577, Japan}
\author{E. Saitoh}
\affiliation{Institute for Materials Research, Tohoku University, Sendai 980-8577, Japan}
\affiliation{WPI Advanced Institute for Materials Research, Tohoku University, Sendai 980-8577, Japan}
\affiliation{Center for Spintronics Research Network, Tohoku University, Sendai 980-8577, Japan}
\affiliation{Advanced Science Research Center, Japan Atomic Energy Agency, Tokai 319-1195, Japan}
\begin{abstract}
The spin Peltier effect (SPE), heat-current generation due to spin-current injection, in various metal (Pt, W, and Au single layers and Pt/Cu bilayer)/ferrimagnetic insulator (yttrium iron garnet: YIG) junction systems has been investigated by means of a lock-in thermography (LIT) method. The SPE is excited by a spin current across the metal/YIG interface, which is generated by applying a charge current to the metallic layer via the spin Hall effect. The LIT method enables the thermal imaging of the SPE free from the Joule-heating contribution. Importantly, we observed spin-current-induced temperature modulation not only in the Pt/YIG and W/YIG systems but also in the Au/YIG and Pt/Cu/YIG systems, excluding the possible contamination by anomalous Ettingshausen effects due to proximity-induced ferromagnetism near the metal/YIG interface. As demonstrated in our previous study, the SPE signals are confined only in the vicinity of the metal/YIG interface; we buttress this conclusion by reducing a spatial blur due to thermal diffusion in an infrared emission layer on the sample surface used for the LIT measurements. We also found that the YIG-thickness dependence of the SPE is similar to that of the spin Seebeck effect measured in the same Pt/YIG sample, implying the reciprocal relation between them.
\end{abstract}
\pacs{72.20.Pa, 72.25.-b, 85.75.-d, 85.80.-b}
\maketitle
%
%
%
%
\section{INTRODUCTION} \label{introduction}
\begin{figure*}[ht]
\begin{center}
\includegraphics{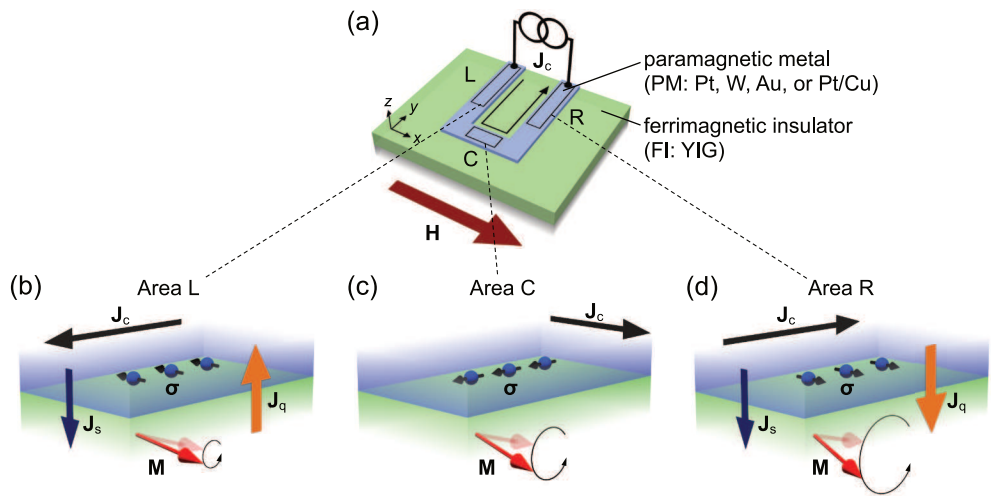}
\caption{(a) Schematic of the PM/FI sample used for the measurements of the SPE. The sample comprises a U-shaped PM (in experiments, Pt, W, or Au single layers or Pt/Cu bilayer) film and an FI (in experiments, YIG). The squares on the PM define the areas L, C, and R. (b)-(d) Schematic illustrations of the SPE induced by the SHE on the areas L (b), C (c), and R (d). ${\bf{J}_{\rm{c}}}$, ${\bf{J}_{\rm{s}}}$, ${\bf{J}_{\rm{q}}}$, ${\bf{M}}$, and ${\bf{H}}$ denote the charge current applied to the PM, spin current with the spin vector ${\bm{\sigma}}$ generated by the SHE in the PM, heat current generated by the SPE near the PM/FI interface, magnetization vector with the magnitude $M$, and magnetic field vector with the magnitude $H$, respectively. When the SHA of the PM and $H$ are positive, ${\bf{M}}$ and ${\bm{\sigma}}$ are antiparallel, perpendicular, and parallel on L, C, and R, respectively. Depending on the relative angle between ${\bf{M}}$ and ${\bm{\sigma}}$, the nonequilibrium energy transport via the interfacial spin exchange generates ${\bf{J}_{\rm{q}}}$, which results in the temperature gradient parallel to ${\bf{J}_{\rm{q}}}$ as schematically shown in our previous paper [\onlinecite{SPE2}].}\label{figure1}
\end{center}
\end{figure*}
A spin current, a flow of the spin angular momentum, plays a crucial role in spintronics [\onlinecite{spintronics1,spintronics2,spintronics3,spintronics4,spintronics5,spincurrent1,spincurrent2}]. Recently, the interaction between spin and heat currents in paramagnet/ferromagnet junction systems has attracted much attention [\onlinecite{spincaloritronics1,spincaloritronics2,spincaloritronics3}]. One important example is the spin Seebeck effect (SSE) [\onlinecite{SSE1,SSE2,SSE3,SSE4,SSE5,SSE6,SSE7,SSE8,SSE9,SSE10,SSE11,SSE12,SSE13,SSE14,SSE15,SSE16,SSE17,SSE18,SSE19,SSE20,SSE21,SSE22,SSE23,SSE24,SSE25,SSE26,SSE27,SSE28,SSE29,SSE30,SSE34,SSE31,SSE32,SSE33}], which refers to the spin-current generation as a result of a heat current. When a temperature gradient is applied to the junction, the heat current induces nonequilibrium dynamics of magnetic moments in the ferromagnet, which injects a spin current into the attached paramagnet [\onlinecite{SSEtheory1,SSEtheory2,SSEtheory3,SSEtheory4,SSEtheory5,SSEtheory6,SSEtheory7,SSEtheory8,SSEtheory9,SSEtheory10,SSEtheory11}]. The spin current is then converted into a charge current by the spin-orbit interaction (SOI) and detected as an electric voltage [\onlinecite{ISHE1,ISHE2,ISHE3}]. Since the SSE appears even in ferromagnetic insulators [\onlinecite{SSE3}], it enables insulator-based thermoelectric generation, which was impossible when only conventional thermoelectric effects are used. Therefore, the SSE has been studied from the viewpoints of fundamental physics as well as thermoelectric application [\onlinecite{spincaloritronics2,application1}].\par
The spin Peltier effect (SPE) [\onlinecite{SPE1,SPE2}] refers to the heat-current generation as a result of a spin current across the paramagnet/ferromagnet junction, which is the Onsager reciprocal of the SSE. The first observation of the SPE was reported by J. Flipse {\it{et al.}} in 2014 using a junction comprising a paramagnetic metal Pt and a ferrimagnetic insulator yttrium iron garnet $\mathrm{Y_3Fe_5O_{12}}$ (YIG) [\onlinecite{SPE1}]. They measured temperature modulation in linear response to the spin-current injection using micro-fabricated thermocouples, and they observed temperature change on the bare YIG surface around the Pt spin injector. Recently, we established another technique for measuring the SPE based on active infrared emission microscopy called lock-in thermography (LIT) [\onlinecite{SPE2,LIT1,LIT2}]. The LIT method enables imaging of the temperature modulation induced by SPEs with high temperature and spatial resolutions ($<0.1$ mK and $\sim$ 6 ${\rm{\mu m}}$, respectively, around room temperature) but requires no micro-fabrication processes, realizing simple and versatile investigation of SPEs. Now, we are ready to carry out systematic studies on SPEs to clarify their nature.\par
In this paper, we report systematic measurements of the SPE using the LIT method using various metal/YIG junction systems. This paper is organized as follows. In Sec. \ref{expconfandproc}, we explain sample configurations and experimental setups for the measurements of the SPE using the LIT method, followed by the details of the experimental procedures. In Sec. \ref{resultanddiscussion}, we show the experimental results and analyses of the SPE in the Pt/YIG, W/YIG, Pt/$\mathrm{Al_2O_3}$/YIG, Au/YIG, and Pt/Cu/YIG junction systems, which confirm that the LIT method enables visualization of the temperature modulation induced by spin currents injected into the YIG layer from the adjacent metal. The Au/YIG and Pt/Cu/YIG samples allow us to remove contributions from conventional thermoelectric effects and to realize pure detection of the SPE. By investigating temperature distribution induced by the SPE in the Pt/YIG sample with a very thin infrared emission layer, we found that the signal is confined only in the vicinity of the Pt/YIG interface within the spatial resolution of the infrared camera ($\sim$ 6 ${\rm{\mu m}}$). In Sec. \ref{reciprocity}, we investigate the YIG-thickness dependence of the SPE and SSE and discuss the reciprocity between them. The Sec. \ref{summary} is devoted to a summary of the present study.\par
\section{EXPERIMENTAL CONFIGURATION AND PROCEDURE} \label{expconfandproc}
\begin{figure*}[ht]
\begin{center}
\includegraphics{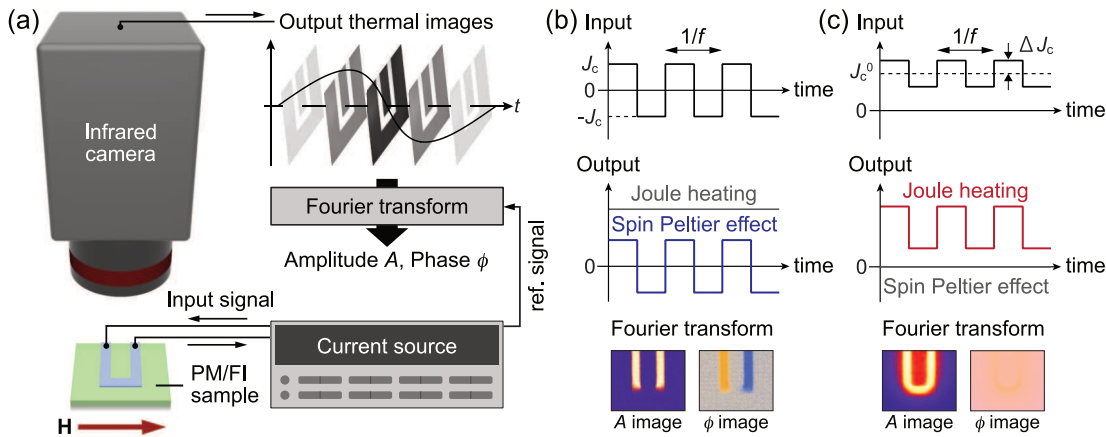}
\caption{(a) Schematic of the lock-in thermography (LIT) measurements. (b) LIT conditions for the SPE measurements. When we apply a square-wave charge current with the amplitude $J_{\rm{c}}$, the frequency $f$, and no DC offset, the SPE-induced temperature modulation ($\propto J_{\rm{c}}$) oscillates with $f$, while the Joule-heating-induced temperature modulation ($\propto J_{\rm{c}}^2$) is constant in time. By extracting the first harmonic response of the thermal images, only the SPE contribution can be detected. (c) LIT conditions for the Joule-heating measurements. When we apply a square-wave current with the amplitude $\Delta J_{\rm{c}}$, the frequency $f$, and the finite DC offset $J_{\rm{c}}^0$, the LIT images are dominated by the temperature modulation induced by the Joule heating because it is much greater than that of the SPE. }\label{figure2}
\end{center}
\end{figure*}
Figure \ref{figure1}(a) shows a schematic illustration of SPEs in a paramagnetic metal (PM)/ferrimagnetic insulator (FI) junction used in the present study. The SPE appears as a result of a spin current generated by the spin Hall effect (SHE) in the PM [\onlinecite{SHE1,SHE2,SHE3,SHE4,SHE5,SHE6}]. When a charge current ${\bf J}_{\rm{c}}$ is applied to the PM with strong SOI, a spin current ${\bf J}_{\rm{s}}$ with a spin vector ${\bm{\sigma}}$ is generated by the SHE, which satisfies the following relation [\onlinecite{SSE27}]:
\begin{equation}
\label{SHE}
{\bf{J}}_{\rm{s}}=\frac{\hbar}{2e}\theta_{\rm{SH}}{\bm{\sigma}}\times{\bf{J}}_{\rm{c}},
\end{equation}
where $e$ ($<0$) and $\theta_{\rm{SH}}$ are the electric charge of an electron and the spin Hall angle (SHA), respectively, and ${\bm{\sigma}}$ is defined as a unit vector. The spin current induces spin accumulation with the spin direction of ${\bf{J}}_{\rm{c}}\times{\bf{n}}$ near the PM/FI interface, where {\bf{n}} is the normal vector of the PM/FI interface plane in the $+z$ direction. The spin accumulation transfers spin angular momentum and energy from electrons in the PM to magnons in the FI via the interfacial spin exchange, i.e., the spin-mixing conductance [\onlinecite{STT}]. This process is proportional to the magnitude of the injected spin current and depends on whether ${\bm{\sigma}}$ in the PM is parallel or antiparallel to the magnetization {\bf{M}} of the FI, which are the characteristics of the spin transfer torque [\onlinecite{STT}]. The nonequilibrium spin and energy transport between electron and magnon systems generates a heat current ${\bf{J}}_{\rm{q}}$ across the PM/FI interface, which satisfies the following symmetry:
\begin{equation}
\label{symm_SPE}
{\bf{J}}_{\rm{q}}\propto({\bm{\sigma}}\cdot{\bf{M}}){\bf{n}}\propto{\bf{J}}_{\rm{c}}\times{\bf{M}}.
\end{equation}
\par
To demonstrate the symmetry of the SPE in a single device, we formed a U-shaped PM layer on the FI [Fig. \ref{figure1}(a)]. In this structure, the relative direction between ${\bf{J}}_{\rm{c}}$ and {\bf{M}} on the PM/FI interface is different among the areas L, R, and C [Figs. \ref{figure1}(b)-(d)]. Owing to the symmetry of the SHE, ${\bm{\sigma}}$ is directed along the -$x$, -$y$, and +$x$ directions at the PM/FI interface on L, R, and C, respectively. When the external magnetic field {\bf{H}} with the magnitude $H$ is applied to the +$x$ direction, ${\bm{\sigma}}$ is antiparallel, parallel, and perpendicular to {\bf{M}} on L, R, and C, respectively. Therefore, the amplitude of the {\bf{M}} fluctuation is suppressed (enhanced) on L (R), while it is not modulated on C, because of the symmetry of the spin transfer torque. Thus, ${\bf{J}}_{\rm{q}}$ is generated along the  +$z$ (-$z$) direction on L (R), while no ${\bf{J}}_{\rm{q}}$ generation appears on C [Eq. (\ref{symm_SPE})].\par
In this study, we measured the SPE in various metal (Pt, W, and Au single layers and Pt/Cu bilayer)/YIG junction systems, where Pt, W, and Au are typical metals showing strong SOI and SHEs [\onlinecite{SHE5}]. The thickness of the Pt, W, and Cu layers (Au layer) is $5\ {\rm{nm}}$ ($10\ {\rm{nm}}$) and the width of the U-shaped structure is $0.2\ {\rm{mm}}$. All the metals were sputtered on a single-crystalline YIG, which was grown on the whole of a single-crystalline $\mathrm{Gd_3Ga_5O_{12}}$ (GGG) substrate by a liquid phase epitaxy method [\onlinecite{LPE}]. Before sputtering the metals, the surface of the YIG was mechanically polished with alumina slurry with the particle diameter of $0.05\ {\rm{\mu m}}$. The thickness of the YIG is $112\ {\rm{\mu m}}$ except for the YIG-thickness-dependent measurements shown in Sec. \ref{reciprocity}. In the YIG-thickness-dependent measurements, we prepared five YIG/GGG substrates with the YIG thickness of $t_{\rm{YIG}}=2.1,\ 5.1,\ 19.6,\ 41.7,\ 109\ {\rm{\mu m}}$ and formed a Pt film with the thickness of $5$ nm and the rectangular shape on each YIG.\par
To detect the temperature modulation induced by the SPE, we employed the LIT method [Fig. \ref{figure2}(a)]. Since typical temperature modulation induced by the SPE is in the order of $1$ mK, it cannot be measured by the conventional steady-state thermography, of which the detection limit is $> 20$ mK. In contrast, the LIT provides the higher temperature resolution of $<0.1\ {\rm{mK}}$ and enables the contact-free measurements of spatial distribution of the SPE signals over a large area [\onlinecite{SPE2}]. The LIT measurement is performed by the following procedures. First, a periodic external perturbation, such as a charge or spin current, is applied to a sample. At the same time, thermal images are measured at a high frame rate ($100$ Hz for our infrared camera). The thermal images are Fourier-transformed at the same frequency as the perturbation. Then, the Fourier-transform amplitude $A$ and phase $\phi$ images of the temperature modulation induced by the perturbation are obtained. The $A$ ($\phi$) image gives the information about the magnitude of the temperature modulation (the sign of the temperature modulation and the time delay due to the thermal diffusion). The amplitude and phase are defined in the range of $A\geq0$ and $0^{\circ}\leq\phi<360^{\circ}$. Here, $\phi=0^{\circ}$ means that  the input perturbation and output temperature change oscillate in the same phase.\par
The SPE measurements using the LIT method is schematically shown in Fig. \ref{figure2}(b). We measured the spatial distribution of infrared radiation emitted from the sample surface with applying a square-wave charge current, whose amplitude and frequency are $J_{\rm{c}}$ and $f$, respectively. We then extracted the first harmonic response of the detected signals. Here, we set $f=5$ Hz except for the frequency-dependent measurements in Sec. \ref{resultanddiscussion}. Importantly, the SPE-induced temperature modulation is proportional to $J_{\rm{c}}$, while the Joule-heating contribution is to $J_{\rm{c}}^2$. Therefore, the Joule heating generated by the square-wave current is constant in time as depicted in Fig. \ref{figure2}(b), which enables us to separate the SPE signals from the Joule-heating signals by using the LIT method. In contrast, as shown in Fig. \ref{figure2}(c), we can measure the temperature modulation induced by the Joule heating by applying a square-wave current with the amplitude of $\Delta J_{\rm{c}}$ and the finite DC offset $J_{\rm{c}}^0$ because the Joule-heating signals appear also in the first harmonic response of the thermal images in this condition, when the SPE signals are much smaller than the Joule-heating contribution.\par
\begin{figure}[ht]
\begin{center}
\includegraphics[width=74mm]{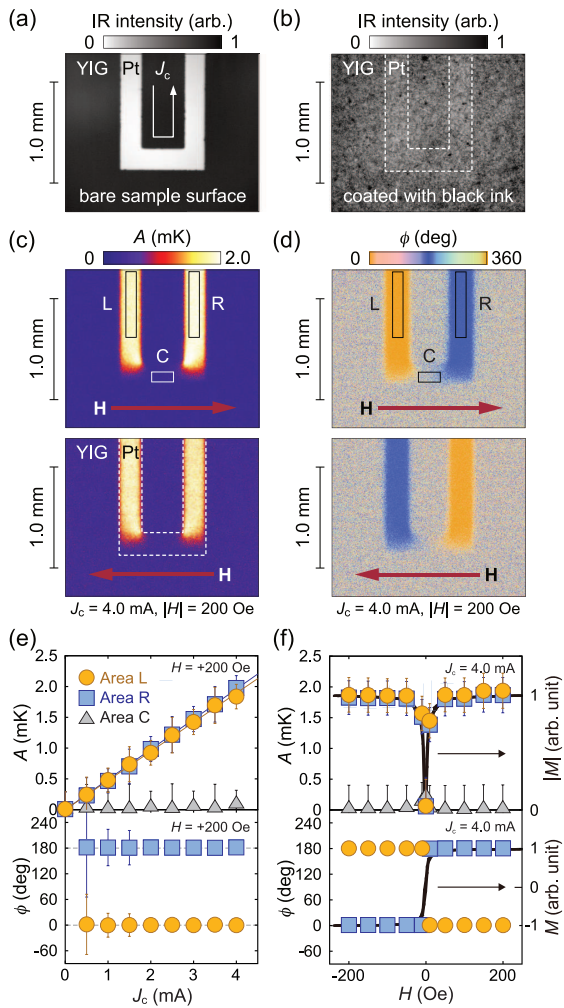}
\caption{(a),(b) Steady-state infrared images of the Pt/YIG sample without (a) and with (b) the black-ink coating at thermal equilibrium. The image in (b) confirms the uniform emissivity of the sample surface. The LIT measurements were performed by using the sample with the black-ink coating except for the experiments in Fig. \ref{figure7}. (c),(d) Lock-in amplitude $A$ (c) and phase $\phi$ (d) images for the Pt/YIG sample at $J_{\rm{c}}=4$ mA. The upper and lower panels show the signals at $H=+200$ and $-200$ Oe, respectively. (e) $J_{\rm{c}}$ dependence of $A$ and $\phi$ on the areas L (yellow circles), R (blue squares), and C (gray triangles) of the Pt/YIG sample at $H=+200$ Oe. (f) $H$ dependence of $A$ and $\phi$ on L, R, and C of the Pt/YIG sample at $J_{\rm{c}}=4$ mA. The $M$-$H$ curve of the YIG is also plotted. The error bars are defined as a standard deviation. The lock-in phase does not converge to a specific value when the signal amplitude is smaller than the sensitivity of the LIT; therefore, the $\phi$ data for C are not shown in (e) and (f).}\label{figure3}
\end{center}
\end{figure}
The temperature of the sample is detected in terms of the emission intensity of the infrared light in the wavelength range of $3$ -- $5$ ${\rm{\mu}}$m in our measurements. Figure \ref{figure3}(a) shows an infrared thermal image of the Pt/YIG sample used in our experiments at room temperature, which was obtained without using the LIT method. The black-and-white contrast in the thermal image comes from the difference in the emissivity between the Pt film and YIG. Significantly, since YIG is transparent in the detectable wavelength range of our infrared camera and its infrared emissivity is almost zero, infrared emission from bare YIG cannot be detected directly (see appendix A) [Note that the black color on the YIG area in Fig. \ref{figure3}(a) is attributed to the infrared emission from the sample stage beneath the sample]. In contrast, the Pt film exhibits significant infrared emissivity ($\sim0.3$ in the wavelength range of $3$ -- $5$ ${\rm{\mu}}$m) owing to the size effect for the electromagnetic response (see appendix B), while the emissivity of metals is very small in general. Therefore, the temperature change on the Pt film can be detected with the infrared camera, although its quantitative estimation is difficult because the emissivity is still less than that of the black body ($=1$). To overcome the low and non-uniform emissivity distribution of the sample, the sample surface was coated with insulating black ink, of which the emissivity is $>0.95$ (except for the results in Fig. \ref{figure7}). The black ink mainly consists of SiZrO$_4$, Cr$_2$O$_3$, and iron-oxide based inorganic pigments, which is commercially available from Japan Sensor Corporation. The thickness of the black ink is $20$ -- $30$ ${\rm{\mu}}$m in our experiments and the infrared light transmittance of the black ink layer is almost zero. Figure \ref{figure3}(b) shows an infrared thermal image of the Pt/YIG sample with the black-ink coating, which confirms high and uniform emissivity of the sample surface.\par
In the LIT experiments, the infrared intensity $I$ detected by the infrared camera needs to be converted into temperature $T$ information. This conversion is done by measuring $T$ dependence of $I$. Since the LIT extracts thermal images oscillating with the same frequency as a periodic external perturbation applied to the sample, the $I$-to-$T$ conversion in the LIT is determined by the differential relation $\Delta T_{1{\rm{f}}}({\bf{r}})=\left.{\rm{d}}T/{\rm{d}}I\right|_{T}({\bf{r}})\Delta I_{1{\rm{f}}}({\bf{r}})$, where $\Delta T_{1{\rm{f}}}({\bf{r}})$ and $\Delta I_{1{\rm{f}}}({\bf{r}})$ denote the lock-in responses of the temperature and infrared radiation intensity at the position ${\bf{r}}$, respectively. In this study, we employed the following five-step calibration method: (1) Measure the $T$ dependence of $I$ in the steady-state condition by using the infrared camera with changing the $T$ value of the sample, (2) Calculate the ${\rm{d}}T/{\rm{d}}I$ function from the obtained $I$-$T$ relation for each pixel, (3) Perform the LIT measurements; measure the first harmonic response of the $I$ images, i.e., $\Delta I_{1{\rm{f}}}$ images, with applying a square-wave charge current to the sample, (4) Determine $T$ values during the LIT measurements for each pixel by using the $I$-$T$ relation and steady-state $I$ images measured in parallel with the $\Delta I_{1{\rm{f}}}$ images, and (5) Convert the $\Delta I_{1{\rm{f}}}$ images into $\Delta T_{1{\rm{f}}}$ images by applying the $\left.{\rm{d}}T/{\rm{d}}I\right|_{T}$ value, obtained from the steps (2) and (4), to each pixel. This calibration method is valid only when the infrared emissivity of the sample surface is high ($\sim$ 1), where infrared light transmitted through and reflected from the sample is negligibly small.\par
\begin{figure*}[ht]
\begin{center}
\includegraphics{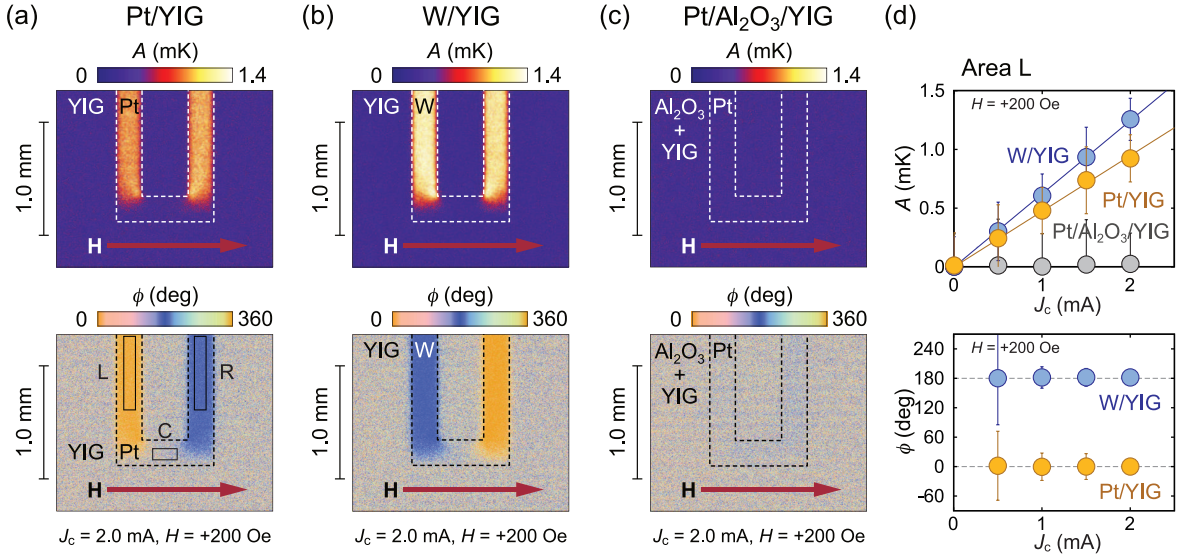}
\caption{(a)-(c) $A$ and $\phi$ images for the Pt/YIG (a), W/YIG (b), and Pt/Al$_2$O$_3$/YIG (c) samples at $J_{\rm{c}}=2$ mA and $H=+200$ Oe. (d) $J_{\rm{c}}$ dependence of $A$ and $\phi$ on L of the Pt/YIG (yellow circles), W/YIG (blue circles), and Pt/Al$_2$O$_3$/YIG (gray circles) samples at $H=+200$ Oe.}\label{figure4}
\end{center}
\end{figure*}
\section{RESULTS AND DISCUSSION} \label{resultanddiscussion}
\subsection{Thermal imaging of the SPE}
First, we show data of current-induced temperature modulation in the Pt/YIG structure. The upper panels in Figs. \ref{figure3}(c) and (d) show the LIT amplitude $A$ and phase $\phi$ images at $J_{\rm{c}}=4$ mA and $H=+200$ Oe (${\bf{H}}||+x$ direction), respectively. The clear temperature modulation was observed on the areas L and R, but not on C, which is consistent with the aforementioned symmetry of the SHE. We found that $\phi=0^{\circ}$ ($\phi=180^{\circ}$) on L (R), showing that the input charge current and output temperature modulation oscillate with the same (opposite) phase on L (R) in the Pt/YIG sample. Since the heat-conduction condition is the same between L and R, the $\phi$ shift of $180^{\circ}$ between L and R is irrelevant to the time delay caused by the thermal diffusion, indicating that the sign of the temperature modulation on the Pt/YIG surface is reversed by reversing the ${\bf{J}}_{\rm{c}}$ direction. Figure \ref{figure3}(e) shows the $J_{\rm{c}}$ dependence of $A$ and $\phi$ at $H=+200$ Oe. The $A$ values are proportional to $J_{\rm{c}}$ and the $\phi$ values remain unchanged with respect to $J_{\rm{c}}$. This result confirms that the observed temperature modulation on L and R of the Pt/YIG sample appears in linear response to the charge current in the Pt layer.\par
We also measured the $H$ dependence of the temperature modulation induced by the charge current by using the same Pt/YIG sample. The sign of the temperature modulation, $\phi$, is reversed by reversing the ${\bf{H}}$ direction [see Fig. \ref{figure3}(d)], indicating that the signal is affected by the ${\bf{M}}$ direction of YIG. As shown in Fig. \ref{figure3}(f), the temperature modulation is an odd function with respect to $H$ and is saturated when the magnetization of YIG is saturated ($|H|>50$ Oe); this is the characteristic feature of the SPE [see Eq. (\ref{symm_SPE})]. The observed clear sign reversal also shows that the contribution from the $H$-independent effects, such as the conventional Peltier effect at the electric contacts, is negligibly small. When $|H|<50$ Oe, the $A$ value rapidly decreases to zero as $|H|$ approaches to zero. This is because the magnitude of {\bf{M}} decreases to zero as shown in Fig. {\ref{figure3}}(f).\par
To verify the origin of the current-induced temperature modulation, we performed some control experiments. Figure \ref{figure4}(b) shows the $A$ and $\phi$ images in the W/YIG sample at $J_{\rm{c}}=4$ mA and $H=+200$ Oe, where the ${\bm{\sigma}}$ direction of the  spin current flowing across the W/YIG interface is opposite to that of the Pt/YIG interface since the sign of the SHA of W and Pt is opposite. On L and R of the W/YIG sample, clear temperature modulation proportional to $J_{\rm{c}}$ was observed [see Figs. \ref{figure4}(b) and (d)]. Importantly, the sign of the temperature modulation in the W/YIG sample was opposite to that of the Pt/YIG sample [compare Figs. \ref{figure4}(a) and (b)]. This sign change is consistent with the sign of the SHA, showing that the signal comes from the spin current generated by the SHE in the PM. We also measured the current-induced temperature modulation in a Pt/$\mathrm{Al_2O_3}$/YIG junction system, where the $1$-${\rm nm}$-thick $\mathrm{Al_2O_3}$ was grown on the YIG by an atomic layer deposition method before forming the Pt layer. Since the spin current generated by the SHE in the Pt layer cannot flow into the YIG layer through the $\mathrm{Al_2O_3}$ layer [\onlinecite{SSE-AlO}], the current-induced temperature modulation should disappear in this structure. In fact, no signal appears in the Pt/$\mathrm{Al_2O_3}$/YIG sample [Figs. \ref{figure4}(c) and (d)], confirming that the signal originates from the injection of the spin current across the PM/YIG interfaces.\par
\begin{figure*}[ht]
\begin{center}
\includegraphics{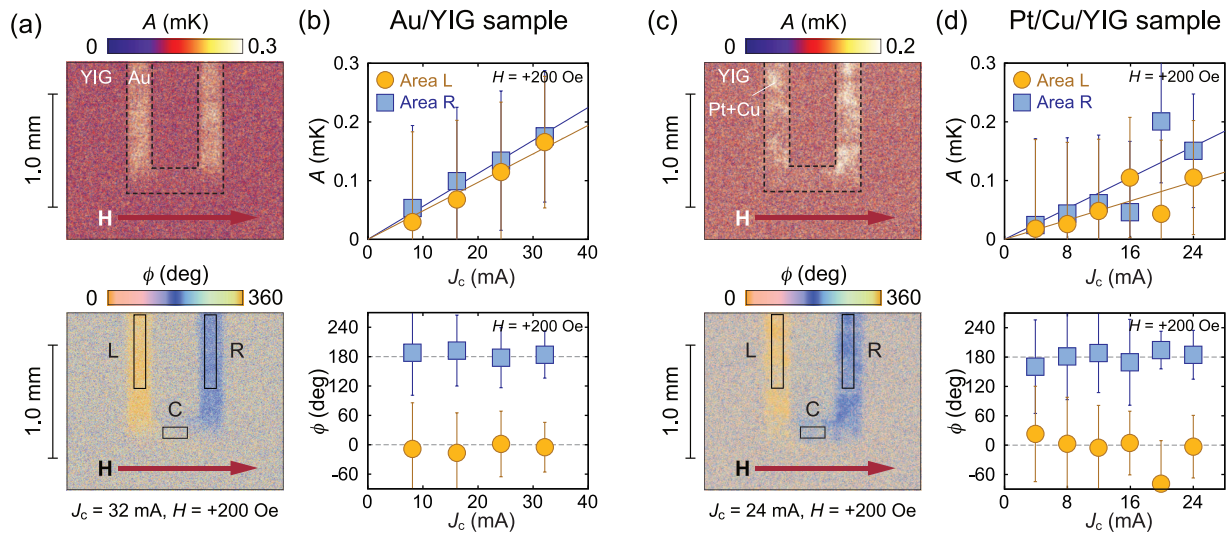}
\caption{(a) $A$ and $\phi$ images for the Au/YIG sample at $J_{\rm{c}}=32$ mA and $H=+200$ Oe. (b) $J_{\rm{c}}$ dependence of $A$ and $\phi$ on L (yellow circles) and R (blue squares) of the Au/YIG sample at $H=+200$ Oe. (c) $A$ and $\phi$ images for the Pt/Cu/YIG sample at $J_{\rm{c}}=24$ mA and $H=+200$ Oe. (d) $J_{\rm{c}}$ dependence of $A$ and $\phi$ on L and R of the Pt/Cu/YIG sample at $H=+200$ Oe.}\label{figure5}
\end{center}
\end{figure*}
\subsection{Separation of the SPE from the anomalous Ettingshausen effect}
The above experiments clearly show that the Pt/YIG and W/YIG samples exhibit the current-induced temperature modulation with the same symmetry as the SPE. However, to complete exclusive establishment of the SPE, we need to separate the SPE from conventional other thermoelectric effects. Although the contribution from the conventional Peltier effect is negligibly small as mentioned before, we have to check the contribution from the Ettingshausen effects in the metal layer, of which symmetry is similar to the SPE [\onlinecite{spincaloritronics2}]. The normal (anomalous) Ettingshausen effect generates a heat current in the direction of the cross product of the applied charge current and external magnetic field (spontaneous magnetization), where the output temperature modulation is proportional to the magnitude of the magnetic field (magnetization). Here, we found that the normal Ettingshausen effect in Pt is negligibly small because $A$ is saturated in the range of $|H|>50$ Oe in the Pt/YIG sample [see Fig. \ref{figure3}(f)]. The anomalous Ettingshausen effect (AEE) in ferromagnetic materials does not exist in our sample, since YIG is a very good electrical insulator. In contrast, in the Pt/YIG system, ferromagnetism may be induced in the Pt layer due to a static magnetic proximity effect in the vicinity of the Pt/YIG interface [\onlinecite{Proximity}], since Pt is near the Stoner ferromagnetic instability [\onlinecite{Instability1,Instability2}]. If the static ferromagnetism in Pt appears and induces temperature modulation by the AEE, it may contaminate the SPE signals. The contribution from the magnetic proximity effect in the Pt/YIG junction was shown to be negligibly small in the SSE experiments [\onlinecite{SSE20,SSE24}], where the possible thermopower due to the proximity-induced anomalous Nernst effect (ANE) is three orders of magnitude smaller than that due to the SSE. When we assume the reciprocal relations in the charge-spin-heat conversion phenomena, the temperature modulation due to the proximity-induced AEE is expected to be much smaller than that due to the SPE in the Pt/YIG sample.\par
To observe the SPE free from the proximity-induced AEE, we performed the LIT measurements using a Au/YIG (Pt/Cu/YIG) junction system, where the Pt layer is replaced with a Au film (a Cu film is inserted between Pt and YIG) to avoid the magnetic proximity effect. Since Au and Cu are typical metals far from the Stoner instability, the Au/YIG and Pt/Cu/YIG samples allow us to demonstrate that the current-induced temperature modulation is irrelevant to the magnetic proximity effect. In the Pt/Cu/YIG sample, while the Cu layer have little ability to generate the spin current due to the small SOI, the spin current generated in the Pt layer passes through the Cu layer and reaches the Cu/YIG interface owing to the large spin diffusion length of Cu [\onlinecite{SHE6}].\par
Figure \ref{figure5}(a) shows the $A$ and $\phi$ images in the Au/YIG sample at $J_{\rm{c}}=32$ mA and $H=+200$ Oe. The current-induced temperature modulation appears also in the Au/YIG sample on L and R, of which the $J_{\rm{c}}$ and $H$ dependences are consistent with those in the Pt/YIG sample. The sign of the temperature modulation in the Au/YIG sample is the same as that in the Pt/YIG sample, consistent with the sign of the SHA of Au [\onlinecite{SHE5,SHE-Au}]. We observed similar signals also in the Pt/Cu/YIG sample as shown in Figs. \ref{figure5}(c) and (d). These results clearly show that the observed temperature-modulation signals in the Au/YIG and Pt/Cu/YIG samples are due purely to the SPE induced by the SHE because of the absence of the proximity-induced AEE at the Au/YIG and Cu/YIG interfaces [\onlinecite{Instability1,Instability2}].\par
In Table \ref{table}, we compare the magnitude of the SPE in the Pt/YIG, W/YIG, Au/YIG, and Pt/Cu/YIG samples in terms of the amplitude $A$ per unit current density $j_{\rm{c}}$ on the PM/YIG interface, where the sign of the temperature modulation on L in the Pt/YIG sample is defined as positive. The magnitude of the SPE in the Pt/YIG and W/YIG samples is much greater than that in the Au/YIG (Pt/Cu/YIG) sample, since the SHA of Pt and W is much larger than that of Au [\onlinecite{SHE5}] (the charge-curret shunting effect and spin diffusion in the Pt/Cu bilayer reduce the spin accumulation at the Cu/YIG interface [\onlinecite{Iguchi_tri}]). The sign of the temperature modulation in all the PM/YIG samples is consistent with the sign of the SHA of the PM [\onlinecite{SHE5}].\par
\begin{table}[htb]
\begin{tabular}{l@{\hspace{0.4cm}}r@{$\ \times\ $}l@{\hspace{0.8cm}}c}\hline\hline
Sample&\multicolumn{2}{l}{$A/j_{\rm{c}}$ (Km$^2$/A)}&sign on L\\\hline
Pt/YIG&$4.7$&$10^{-13}$&positive\\
W/YIG&$6.2$&$10^{-13}$&negative\\
Au/YIG&$5$&$10^{-15}$&positive\\
Pt/Cu/YIG&$5$&$10^{-15}$&positive\\\hline\hline
\end{tabular}
\caption{Amplitude of the SPE signal $A$ per unit current density $j_{\rm{c}}$ and sign of the temperature modulation on L  for the Pt/YIG, W/YIG, Au/YIG, and Pt/Cu/YIG samples.}
\label{table}
\end{table}
\begin{figure*}[ht]
\begin{center}
\includegraphics{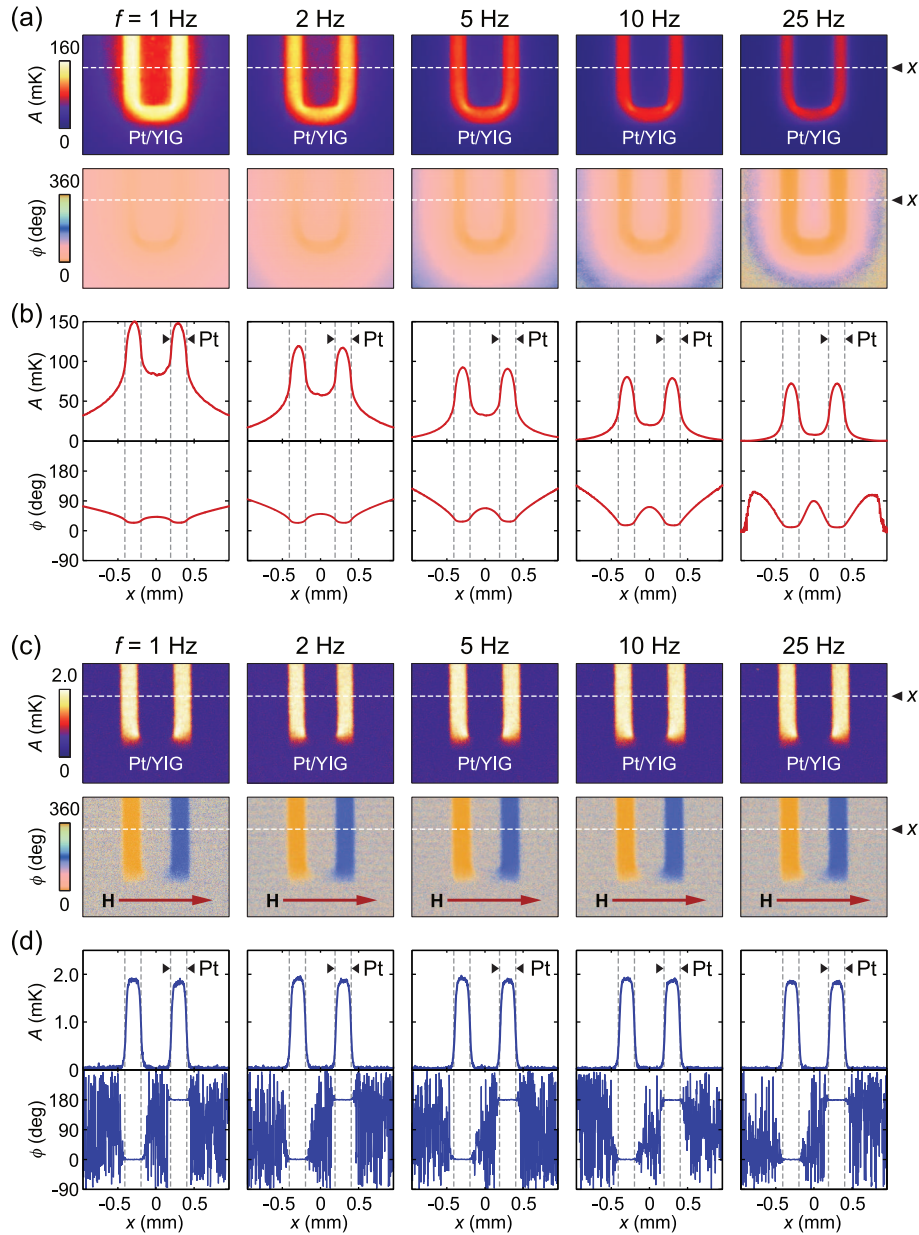}
\caption{(a) $f$ dependence of $A$ and $\phi$ images for the Pt/YIG sample in the Joule-heating condition at $J_{\rm{c}}^0=4$ mA and $\Delta J_{\rm{c}}=0.4$ mA. (b) One-dimensional $A$ and $\phi$ profiles along the $x$ direction across L and R of the Pt/YIG sample in the Joule-heating condition. (c) $f$ dependence of $A$ and $\phi$ images for the Pt/YIG sample in the SPE condition at $J_{\rm{c}}=4$ mA and $H=+200$ Oe. (d) One-dimensional $A$ and $\phi$ profiles along the $x$ direction across L and R of the Pt/YIG sample in the SPE condition. The $\phi$ profiles are noisy because the $\phi$ value dose not converge to a specific value when the $A$ value is smaller than the sensitivity of the LIT.}\label{figure6}
\end{center}
\end{figure*}
\subsection{Spatial distribution of temperature modulation induced by the SPE and Joule heating}
Now we focus on the temperature distribution induced by the SPE. As already demonstrated in Figs. \ref{figure3}, \ref{figure4}, and \ref{figure5}, the SPE signals appear near the PM/YIG interfaces at $f=5$ Hz. However, the LIT images do not reflect the steady-state temperature distribution if the oscillation period of the input signal is shorter than the thermalization time scale of the sample. Therefore, to investigate the temperature distribution induced by the SPE, we have to measure the $f$ dependence of the LIT thermal images; the images at lower $f$ values are closer to the temperature distribution in the steady-state condition.\par
First, to provide typical $f$ dependence of temperature distribution, we measured the Joule heating generated by the charge current in the Pt layer of the Pt/YIG sample by using the measurement condition shown in Fig. \ref{figure2}(c). As shown in the $A$ and $\phi$ images in Fig. \ref{figure6}(a), where $\Delta J_{\rm{c}}=0.4$ mA and $J_{\rm{c}}^0=4.0$ mA, the Joule heating increases the temperature of the Pt layer irrespective to the ${\bf{J}}_{\rm{c}}$ direction and the magnitude of the temperature modulation gradually decreases with the distance from the Pt layer due to the thermal diffusion. We found that the temperature profile on the sample surface strongly depends on the $f$ value [see Fig. \ref{figure6}(b)]. With decreasing $f$, the magnitude of the temperature modulation due to the Joule heating increases and the temperature distribution is broadened in the lateral directions by thermal diffusion; this $f$ dependence of the LIT images is the typical behavior of the temperature change generated from a heating or cooling source.\par
The $f$ dependence of the SPE-induced temperature profile is in sharp contrast to that of the Joule heating. Surprisingly, the temperature modulation profile due to the SPE was found to be independent of the $f$ values as shown in Fig. \ref{figure6}(d). These results indicate that the temperature modulation induced by the SPE immediately reaches the steady state and that the temperature modulation is confined near the Pt/YIG interface even in the steady-state condition. This behavior is quite different from the thermal diffusion expected from conventional heat sources.\par
\begin{figure*}[ht]
\begin{center}
\includegraphics{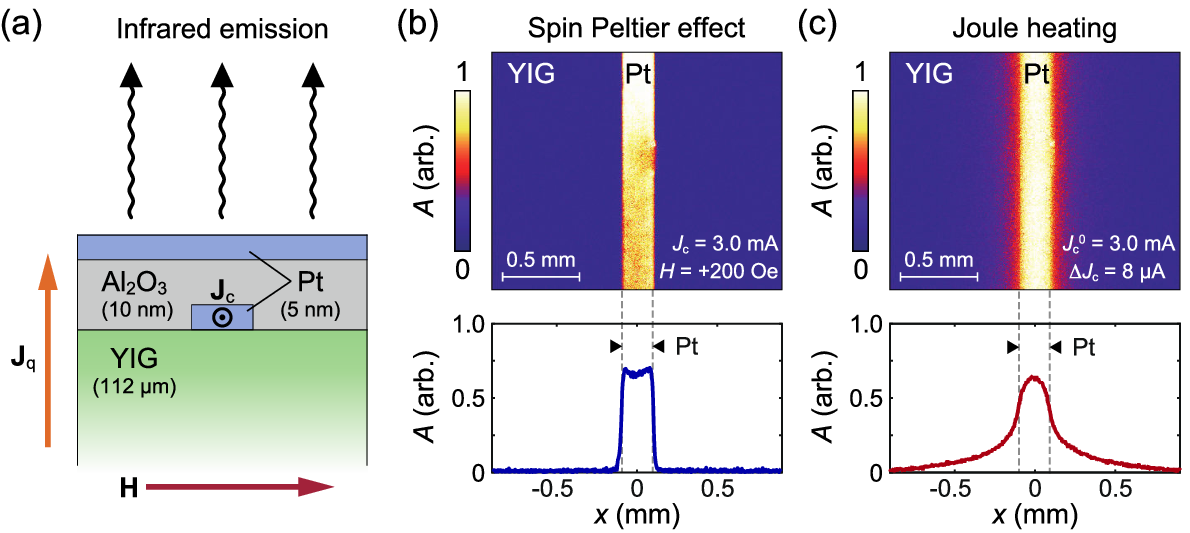}
\caption{(a) Schematic of the cross section of the Pt/Al$_2$O$_3$/Pt/YIG sample used in the measurements without the black-ink coating. We fabricated the Pt/Al$_2$O$_3$/Pt/YIG structure by the following procedures. First, a Pt wire with a thickness of $5$ nm and a width of $200$ $\mu$m was sputtered on the YIG surface. Next, the $10$ nm-thick Al$_2$O$_3$ was grown on the sample by the atomic layer deposition method. Then, the Pt pad with a thickness of $5$ nm was sputtered on the whole surface of the sample. The top Pt layer acts as an infrared-emission layer because of a finite emissivity (see Appendix B for more details). The Al$_2$O$_3$ is a separation layer for insulating the two Pt layers. (b) $A$ image and one-dimensional $A$ profile for the Pt/Al$_2$O$_3$/Pt/YIG sample in the SPE condition at $J_{\rm{c}}=3.0$ mA and $H=+200$ Oe. (c) $A$ image and one-dimensional $A$ profile for the Pt/Al$_2$O$_3$/Pt/YIG sample in the Joule-heating condition at $J^0_{\rm{c}}=3.0$ mA and $\Delta J_{\rm{c}}=8\ \mu{\rm{A}}$.}\label{figure7}
\end{center}
\end{figure*}
The above experiments show that the SPE signal in the Pt/YIG sample is confined near the Pt/YIG interface in the steady-state condition. However, the temperature distribution still contains the thermal diffusion in the black-ink infrared emission layer on the sample surface, of which the thickness is $20$ -- $30$ ${\rm{\mu m}}$. The black ink layer prevents us to observe a bare temperature profile generated by the SPE since the thermal diffusion in the black ink blurs the temperature profile and the spatial resolution is reduced to the values comparable to the black-ink thickness. To further buttress our conclusion that the SPE signal is confined only in the vicinity of the PM/YIG interface, we measured the spatial distribution of the temperature modulation with reducing the spatial blur due to the thermal diffusion in the thick black ink. This is realized by replacing the black ink with a much thinner emission layer; notable is that a $5$-nm-thick Pt film has the finite emissivity of $\sim0.3$ in the detectable wavelength range ($3$ -- $5$ $\mu$m) of our LIT system (see Appendix B), which enables the detection of bare temperature distribution without the spatial blur in the infrared emission layer. While it is difficult to estimate the actual temperature using such an infrared emission layer with low emissivity, the spatial distribution of the temperature is measurable and meaningful as long as the sample has the uniform emissivity. To do this, we measured the SPE-induced temperature profile using a Pt/Al$_2$O$_3$/Pt/YIG sample, where the top Pt film acts as an infrared emission layer and the Al$_2$O$_3$ layer on the Pt/YIG structure is an insulating layer for separating two Pt layers [see Fig. \ref{figure7}(a)]. We found that the temperature distribution induced by the Joule heating in the Pt/Al$_2$O$_3$/Pt/YIG sample is similar to that in the black-ink/Pt/YIG sample [see Fig. \ref{figure7}(c)], confirming that the top Pt film acts as the infrared emission layer. Using the Pt/Al$_2$O$_3$/Pt/YIG sample, we measured the temperature distribution induced by the SPE. Figure \ref{figure7}(b) shows that the SPE-induced temperature modulation is confined near the Pt/YIG interface within the range of the spatial resolution of several $\mu$m.\par
The anomalous temperature distribution induced by the SPE can be explained by assuming the presence of a dipolar heat source, a pair of positive and negative heat-source components with no net heat amount, near the PM/YIG interface, as demonstrated by our numerical calculations shown in Ref. [\onlinecite{SPE2}]. The above experiments imply that the size of the dipolar heat source is less than the spatial resolution of our infrared camera ($\sim6\ {\rm{\mu m}}$). However, the size of the dipolar heat source, the length scale of the SPE, remains undetermined, which may be obtained by detailed and systematic thickness dependent measurements of the SPE.\par
The LIT method reveals the SPE-induced temperature modulation, allowing us to estimate the actual magnitude of the SPE signals. Since the temperature modulation is confined near the PM/YIG interface, the SPE signals should be estimated on the PM/YIG interface. We found that the magnitude of the SPE signals in our Pt/YIG sample, shown in Table \ref{table}, is $57$ times greater than that reported by Flipse {\it{et al.}} [\onlinecite{SPE1}]. The underestimate in Ref. [\onlinecite{SPE1}] may be attributed to the fact that the temperature modulation on the bare YIG surface near the Pt/YIG interface was detected using thermocouples in the previous experiments.\par
\section{Comparison of thickness dependence between SPE and SSE}\label{reciprocity}
\begin{figure*}[ht]
\begin{center}
\includegraphics{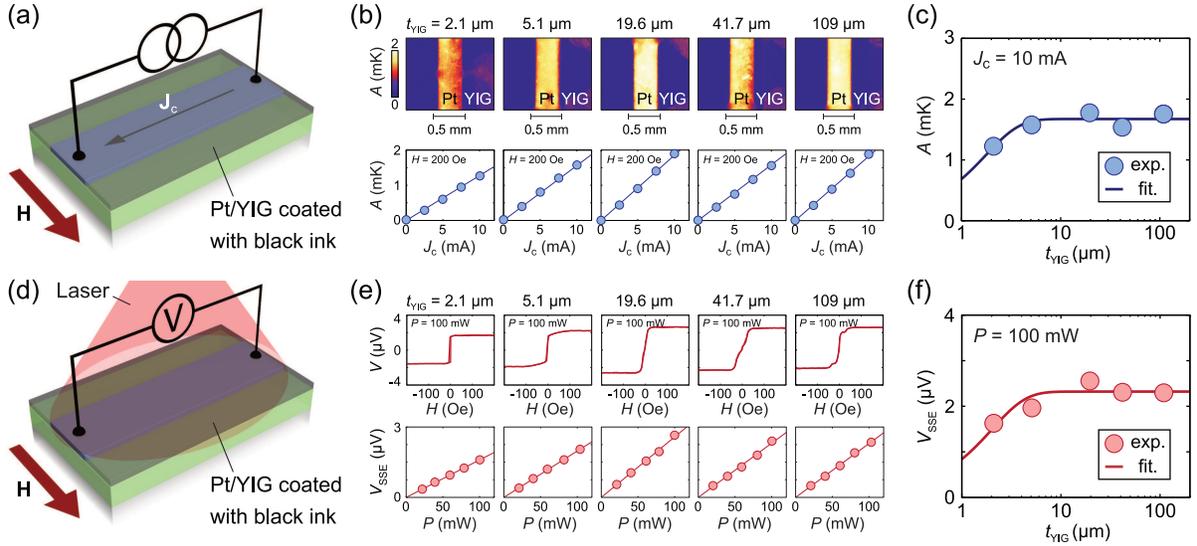}
\caption{(a) Schematic of the Pt/YIG sample for measuring the YIG-thickness $t_{\rm{YIG}}$ dependence of the SPE. (b) $A$ images at $J_{\rm{c}}=10$ mA and $H=+200$ Oe (upper figures) and $J_{\rm{c}}$ dependence of $A$ at $H=+200$ Oe (lower figures) for the Pt/YIG samples with $t_{\rm{YIG}}=2.1,\ 5.1,\ 19.6,\ 41.7,$ and $109\ {\rm{\mu m}}$. (c) $t_{\rm{YIG}}$ dependence of $A$ at $J_{\rm{c}}=10$ mA and $H=+200$ Oe for the Pt/YIG samples. (d) A schematic of the Pt/YIG sample for measuring the $t_{\rm{YIG}}$ dependence of the SSE. The SSE was measured by the laser heating method, where a laser with the wavelength of $670$ nm is applied to the top of the sample uniformly. The black-ink layer used as the infrared emission layer in the SPE measurements acts also as an absorption layer for the laser light in the SSE measurements, resulting in the generation of a heat current across the Pt/YIG interface. $V$ and $P$ denote the voltage induced by the laser heating and the laser power, respectively. (e) $H$ dependence of $V$ at $P=100$ mW (upper figures) and $P$ dependence of the SSE voltage (lower figures) for the Pt/YIG samples. We define the SSE voltage as $V_{\rm{SSE}}=$[$V$($+100$ Oe)-$V$($-100$ Oe)]$/2$. (f) $t_{\rm{YIG}}$ dependence of $V_{\rm{SSE}}$ at $P=100$ mW in the Pt/YIG samples. Solid lines in (c) and (f) show the fitting results obtained from Eq. (\ref{fitting}).}\label{figure8}
\end{center}
\end{figure*}
To discuss physics of spin-heat current conversion phenomena, it is beneficial to investigate the reciprocity between the SPE and SSE. Although we have established the versatile technique for measuring the SPE based on the LIT method, the rigorous verification of the reciprocity between the two effects is still difficult because it requires accurate information about spin transport properties and temperature distribution across the PM/YIG interface. Nevertheless, the relative comparison of the YIG thickness $t_{\rm{YIG}}$ dependence between the SPE and SSE is meaningful. To do this, we measured the SPE and SSE in the Pt/YIG samples with various values of $t_{\rm{YIG}}$ without changing thermal conditions around the samples. The samples with different YIG thickness were coated with the black ink for the SPE measurements. To measure the SSE in the same condition, we adopted the laser heating method [see Fig. \ref{figure8}(d)] [\onlinecite{SSE9,SSE19}], which enables the SSE measurements without attaching a heater on the sample surface.\par
In Fig. \ref{figure8}(b), we show the $t_{\rm{YIG}}$ dependence of the $A$ images and $A$-$J_{\rm{c}}$ relation obtained from the SPE measurements in the Pt/YIG samples. All the samples with different $t_{\rm{YIG}}$ exhibit clear SPE signals. As shown in Fig. \ref{figure8}(c), the temperature modulation due to the SPE increases gradually with increasing $t_{\rm{YIG}}$ and it is saturated when $t_{\rm{YIG}}>10$ ${\rm{\mu m}}$.\par
The upper panels of Fig. \ref{figure8}(e) show the $H$ dependence of the voltage $V$ between the ends of the Pt layer obtained from the SSE measurements in the Pt/YIG samples for various values of $t_{\rm{YIG}}$ at the laser power $P$ of $100$ mW. We successfully observed clear voltage signals of which the sign is reversed in response to the magnetization reversal of YIG. We confirmed that the $V_{\rm{SSE}}$ \{=[$V$($+100$ Oe)-$V$($-100$ Oe)]/2\} values are proportional to $P$ in all the Pt/YIG samples as plotted in the lower panels in Fig. \ref{figure8}(e), which is the characteristic of the SSE. The $t_{\rm{YIG}}$ dependence of $V_{\rm{SSE}}$ at $P$ = $100$ mW is summarized in Fig. \ref{figure8}(f). The SSE signal also increases gradually with increasing $t_{\rm{YIG}}$ and it is saturated when $t_{\rm{YIG}}>10$ ${\rm{\mu m}}$, a behavior similar to that of the SPE.\par
To quantitatively compare the $t_{\rm{YIG}}$ dependence of the SPE and SSE signals, we analyzed the above results in terms of the magnon diffusion length $l_{\rm{m}}$. As reported in the previous studies on the SSE, magnon propagation in the FI plays a crucial role in the SSE [\onlinecite{SSEtheory11}]. The $t_{\rm{YIG}}$ dependence of $V_{\rm{SSE}}$ can be fitted with the following phenomenological equation:
\begin{equation}
\label{fitting}
V_{\rm{SSE}}\propto \frac{\cosh(t_{\rm{YIG}}/l_{\rm{m}})-1}{\sinh(t_{\rm{YIG}}/l_{\rm{m}})},
\end{equation}
where $l_{\rm{m}}$ is the adjustable parameter. The solid line in Fig. \ref{figure8}(f) shows the fitting result; the obtained $l_{\rm{m}}$ value is $1.2$ ${\rm{\mu m}}$, consistent with the results in Refs. [\onlinecite{SSE28,SSE31}]. We apply the same fitting analysis to the SPE signal as plotted in Fig. \ref{figure8}(c). The fitted value of $l_{\rm{m}}$ for the SPE is $1.3$ ${\rm{\mu m}}$, which is comparable to that of the SSE. The similarity in the YIG-thickness dependence of the SPE and SSE suggests that both the effects are governed by the same length scale, implying the reciprocity between them [\onlinecite{SPE1}]. However, we note again that it is difficult to quantitatively discuss the reciprocity between the SSE and SPE because of the difference in the temperature profile in the real experimental setup. The temperature gradient in the SSE is applied to the entire sample, while the temperature modulation in the SPE is confined near the Pt/YIG interface. The difference in the temperature profile between the SPE and SSE makes it difficult to discuss the length scale of spin currents. To investigate the reciprocal relation in more detail, the accurate information about the size of the heat sources in the SPE and the magnon-spectral non-uniform nature of the SPE and SSE [\onlinecite{SSE28}] are necessary.\par
\section{SUMMARY} \label{summary}
In this paper, we measured the spin Peltier effect (SPE), the temperature modulation due to spin-current injection, in the Pt/$\mathrm{Y_3Fe_5O_{12}}$ (YIG), W/YIG, Pt/$\mathrm{Al_2O_3}$/YIG, Au/YIG, and Pt/Cu/YIG samples by means of the lock-in thermography (LIT) technique. The LIT method enables the thermal imaging of the SPE free from the Joule-heating contribution. The current-induced temperature modulation in the Pt/YIG and W/YIG samples satisfies the symmetry of the SPE driven by the spin Hall effect. We observed spin-current-induced temperature modulation also in the Au/YIG and Pt/Cu/YIG systems, confirming that the signals appear even in the absence of the anomalous Ettingshausen effect due to proximity-induced ferromagnetism near the PM/YIG interface. We also measured the spatial distribution of the temperature modulation induced by the SPE and Joule heating in the Pt/YIG sample. It was found that the SPE-induced temperature modulation is confined near the Pt/YIG interface even when we reduce the spatial blur due to the thermal diffusion in an infrared emission layer on the sample surface, while the Joule-heating-induced temperature modulation is broadened from the heat source. Finally, we discussed the reciprocity between the SPE and the spin Seebeck effect (SSE) by comparing the YIG-thickness dependence of these phenomena. We found that the YIG-thickness dependence of the SPE is similar to that of the SSE measured in the same Pt/YIG samples, implying that both the effects are governed by the same length scale. We anticipate that the systematic SPE data reported here will be useful for clarifying the mechanism of the SPE and for developing theories of the spin-heat conversion phenomena.
\section*{ACKNOWLEDGMENTS}
%
%
%
The authors thank J. Shiomi, A. Miura, T. Oyake, M. Matsuo, Y. Ohnuma, T. Kikkawa, D. Hirobe, T. Taniguchi, B. J. van Wees, G. E. W. Bauer, and S. Maekawa for valuable discussions. This work was supported by PRESTO ``Phase Interfaces for Highly Efficient Energy Utilization'' (JPMJPR12C1) and ERATO ``Spin Quantum Rectification Project'' (JPMJER1402) from JST, Japan, Grant-in-Aid for Scientific Research (A) (JP15H02012) and Grant-in-Aid for Scientific Research on Innovative Area ``Nano Spin Conversion Science'' (JP26103005) from JSPS KAKENHI, Japan, NEC Corporation, the Noguchi Institute, and E-IMR, Tohoku University. S.D. is supported by JSPS through a research fellowship for young scientists (JP16J02422). T.H. is supported by GP-Spin at Tohoku University. 
%
%
%
%
%
\section*{APPENDIX A: MEASUREMENTS OF EMISSIVITY OF YIG} \label{appendixa}
Emissivity is important in infrared emission spectroscopy. In emission spectroscopy measurements, we measure intensity of light emitted from materials. When the material is a black body, which ideally absorbs all incident light, the emission intensity obeys Planck's law [\onlinecite{Kirchhoff}]. However, the intensity of the light emitted from real materials is lower than that expected form Planck's law. The ratio of the emission intensity between the material and black body is called emissivity. In the thermal equilibrium, the emissivity is equal to the absorptivity, which is known as Kirchhoff's law [\onlinecite{Kirchhoff,emissivity1}]:
\begin{equation}
\label{equation-kirchhoff}
\epsilon=\frac{1}{2}\sum_{\gamma={\rm{TE,TM}}}\left(1-|R_{\gamma}|^2-|T_{\gamma}|^2\right),
\end{equation}
where $R_{\gamma}$ and $T_{\gamma}$ are the reflection and transmission coefficients, respectively. TE and TM denote transverse electric and magnetic waves, respectively. Kirchhoff's law is a natural consequence of the energy conservation law. When the incident light interacts with elementally excitations (e.g., phonons, electrons, etc.) in the material, a part of the light is absorbed by the material. The absorbed light is then emitted from the material in a relaxation process of the elementally excitations. In the thermal equilibrium, the energy of the absorbed and emitted light must be the same; this is Kirchhoff's law. Therefore, the emissivity is evaluated by measuring the absorptivity.\par
To estimate the emissivity of YIG, we carried out Fourier transform infrared spectroscopy (FTIR) measurements. Here, we measured the wavelength $\lambda$ dependence of the infrared light transmittance and reflectance of a $106$-$\mu$m-thick YIG slab without a substrate. The YIG slab was illuminated with the infrared light at normal incidence, where $\lambda$ was swept from $2$ to $18$ ${\rm{\mu}}$m. Then, the emissivity of the YIG slab was estimated by using Eq. (\ref{equation-kirchhoff}). Note that the emissivity obtained here is a directional emissivity $\epsilon_{\rm{d}}$ for the direction normal to the YIG surface, while the emissivity depends on the direction of wave vectors in general.\par
\begin{figure}[t]
\begin{center}
\includegraphics{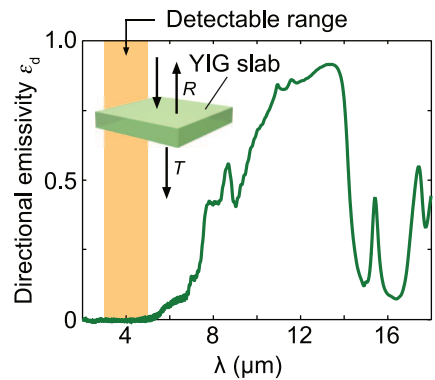}
\caption{Wavelength $\lambda$ dependence of the directional emissivity $\epsilon_{\rm{d}}$ of the YIG. To estimate $\epsilon_{\rm{d}}$ from Kirchhoff's low [Eq. (\ref{equation-kirchhoff})], we measured the reflection $R$ and transmission $T$ coefficients of the YIG with the thickness of 106 ${\rm{\mu}}$m by Fourier transform infrared spectroscopy. The orange area shows detectable wavelength range of our infrared camera.}\label{figure9}
\end{center}
\end{figure}

Figure \ref{figure9} shows the $\lambda$ dependence of $\epsilon_{\rm{d}}$ for the YIG. Importantly, $\epsilon_{\rm{d}}$ of YIG is almost zero when $\lambda<5$ $\mu$m. This is because there is no interaction between YIG and light in this wavelength range. Since the detectable wavelength range of the infrared light in our thermography system is $3$ -- $5$ $\mu$m, one cannot measure the temperature of the bare YIG surface. To measure the temperature of YIG, an infrared emission layer, such as the black-ink layer, has to be formed on the YIG surface. In contrast, in the range of $\lambda>5$ $\mu$m, YIG shows non-zero emissivity because of an interaction between phonons in YIG and the light. When an infrared detector covers this wavelength range (e.g., a microbolometer infrared detector), the infrared emission from the YIG itself is detectable even in the absence of the emission layer.\par
\section*{APPENDIX B: NUMERICAL CALCULATION OF EMISSIVITY OF Pt}
\begin{figure}[ht]
\begin{center}
\includegraphics{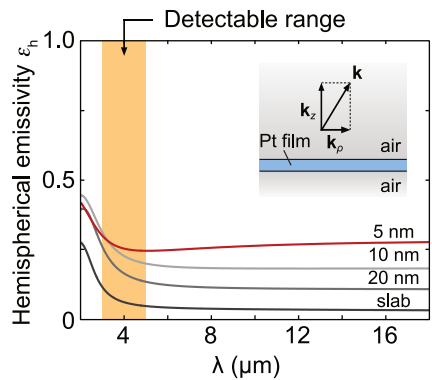}
\caption{Wavelength $\lambda$ dependence of the hemispherical emissivity $\epsilon_{\rm{h}}$ of the Pt films with the thickness of $5$, $10$, and $20$ nm and Pt slab. The inset shows a model used for calculating $\epsilon_{\rm{h}}$. ${\bf{k}}$, ${\bf{k}}_z$, and ${\bf{k}}_{\rho}$ denote the wave vector of the emitted light, projected vector of the wave vector onto the $z$ axis, and projected vector of the wave vector parallel to the surface of the Pt, respectively.}\label{figure10}
\end{center}
\end{figure}
To understand the infrared emission from thin Pt films used for the experiments shown in Fig. \ref{figure7}, we calculate the emissivity of Pt by taking into account the size effect [\onlinecite{emissivity1}]. Emission of light from a thin film is characterized by a hemispherical emissivity $\epsilon_{\rm{h}}$ [\onlinecite{emissivity1}], which is an emissivity for the perpendicular-to-plane component of emitted light. The emitted light is detected as an energy flow propagating perpendicular to the sample surface in experiments. Since the energy flow depends on a wave vector ${\bf{k}}$, the ${\bf{k}}$ dependence of emissivity needs to be taken into consideration. Here, $\epsilon_{\rm{h}}$ is defined as follows [\onlinecite{emissivity1,emissivity2,emissivity3}]:
\begin{equation}
\label{equation-kirchhoff2}
\epsilon_{\rm{h}}=\frac{1}{k_0^2}\int^{k_0}_0k_{\rho}\mathrm{d}k_{\rho}\sum_{\gamma={\rm{TE,TM}}}\left(1-|R_{\gamma}|^2-|T_{\gamma}|^2\right),
\end{equation}
where $k_0$ and $k_{\rho}$ are the magnitude of the wave vector of the light and projection component of the wave vector onto the surface of the Pt. When there is no ${\bf{k}}$ dependence of $R$ and $T$, the hemispherical emissivity is identical to the directional emissivity defined in Eq. (\ref{equation-kirchhoff}).\par
To evaluate $\epsilon_{\rm{h}}$ of Pt films, we performed numerical calculations. The right hand side of Eq. (\ref{equation-kirchhoff2}) was calculated by using the ${\bf{k}}$ dependence of $R$ and $T$ obtained from dielectric constants of the Pt films with the size effect, i.e., the multiple reflection and interference of electromagnetic waves at the top and bottom planes of the films [\onlinecite{material}]. The calculations were performed for the Pt films with various thicknesses of $5$, $10$, and $20$ nm. For comparison, we also calculated $\epsilon_{\rm{h}}$ for a Pt slab without the size effect.\par
In Fig. \ref{figure10}, we show the $\lambda$ dependence of $\epsilon_{\rm{h}}$ of the Pt films with different thicknesses and the Pt slab. In the case of the Pt slab, most of the incident light is reflected at the surface, resulting in small absorption and emission of the light. In contrast, We found that the thinner Pt film shows the higher $\epsilon_{\rm{h}}$ due to the size effect. This behavior indicates that the thinner Pt film has larger light absorption. When $\lambda<5$ ${\rm{\mu m}}$, $\epsilon_{\rm{h}}$ increases with decreasing $\lambda$. This behavior comes from an interband transition in the electronic structure of Pt [\onlinecite{material}]; the resonant interaction between electrons in Pt and the light provides the large absorption and emission. Therefore, thin Pt films can be used as infrared emission layers as demonstrated in Fig. \ref{figure7}.\par

\end{document}